\documentclass[twocolumn,prl]{revtex4}
\usepackage{graphics,graphicx}

\begin{document}
\title{Altermagnetism in NiSi and antiferromagnetic candidate materials with non-collinear spins}
\author{Deepak K. Singh$^{1,2,*}$}
\author{Sang-Wook Cheong$^3$}
\author{Jiasen Guo$^{4}$}
\affiliation{$^{1}$Department of Physics and Astronomy, University of Missouri, Columbia, MO, USA}
\affiliation{$^{2}$MU Materials Science and Engineering Institute, Columbia, MO, USA}
\affiliation{$^{3}$Department of Physics and Astronomy, Rutgers, The State University of New Jersey, NJ}
\affiliation{$^{4}$Oak Ridge National Laboratory, Oak Ridge, TN, USA}
\affiliation{$^{*}$email: singhdk@missouri.edu}

\begin{abstract}

\textbf{Recently, a new class of magnetic phenomenon, called altermagnetism, was proposed where the underlying spin configuration resembles antiferromagnetic structure, but the system violates \textbf{PT} (PT: Parity times Time reversal) symmetry due to the alternation of crystalline symmetry across magnetic ions. Although the original idea was proposed for the collinear spin structure, a recent report by Cheong et al. has suggested that antiferromagnetic materials with non-collinear spin structure and local alternation of crystalline arrangement can also manifest altermagnetism. Besides breaking the \textbf{PT} symmetry, altermagnetic compounds are also expected to exhibit anomalous Hall effects of odd orders. Here, we discuss possible candidates in this regard. One example is nickel monosilicide, which was recently shown to exhibit high temperature antiferromagnetism with non-collinear spin structure. It fulfills both criteria of breaking the \textbf{PT} symmetry and manifesting nonlinear anomalous Hall effect. In addition to NiSi, we also discuss other potential antiferromagnetic materials with non-collinear spin configuration for the exploration of altermagnetic states.}
 
\end{abstract}

 \maketitle

\textbf{Introduction}

Conventionally, magnetism is divided into the two main categories: ferromagnetism and antiferromagnetism. There is also ferrimagnetism and weak ferromagnetism, which basically relates to ferromagnetism. While ferromagnetism is described by the net magnetization, primarily arising from the parallel alignment of spins, an antiferromagnetic material has basically no net magnetization.\cite{Stohr} In the latter case, spin is flipped on the nearest neighboring sites. However, both ferromagnetic and antiferromagnetic systems fulfill important symmetry requirements e.g. translation, rotation, inversion or parity and time reversal.\cite{Skomski} Now, imagining a scenario where the spin structure exhibits primarily antiferromagnetic characteristic, but the lattice symmetry breaks the inversion (parity) symmetry i.e. the crystalline arrangement around the magnetic ion is not exactly same to the neighboring ion. Rather, a rotational operation is required to become equivalent to each other i.e. the same crystalline arrangement on neighboring sites cannot be achieved by mere translation operation. This new type of antiferromagnetic system is called altermagnetism.\cite{Mazin,Smejkal,Smejkal2,Mazin2,Mccarthy}

Although, the underlying physics, describing altermagnetism, has been there for long time, it is only recently that this symmetry-centered phenomenon has gained prominence. Since the inception, a lot of research activities have taken place to identify potential candidate materials for altermagnetism. But most efforts are focused on the antiferromagnetic systems with collinear magnetic structure.\cite{Mazin2,Lee2,Osumi,Lovesey,Aoyama,Reimers,Zhou,Yuan,Libor2,Cui,Amar} That’s because the theoretical hypothesis behind altermagnetism utilizes the collinear spin structure framework.\cite{Feng,Roig,Leeb,Libor,Krempasky,Turek,Fernandes,Helena,Carmine,Fedchenko,Hayami} More recently, Cheong et al. have proposed an extensive group theory-based classification to identify altermagnets with non-collinear spin configurations.\cite{Sang} According to the report, there are two types of altermagnetic states: type-I with net non-zero magnetization and type-ll with fully compensated antiferromagnetic unit cell. Type-I altermagnet involves those materials that are primarily antiferromagnetic, but the magnetic unit cell is not fully compensated. Rather, a small ferromagnetic component accompanies the antiferromagnetic order parameter, most likely arising from the spin-orbital interaction. However, in both cases, parity (inversion) and time reversal symmetry (jointly termed as \textbf{PT} symmetry) are broken and only rotational symmetry operation can map one spin-sublattice onto the other. As Mazin et al. have pointed out, due to this peculiar symmetry constraint, the electronic spectra of the two spin-sublattices in an altermagnet are different. Consequently, the spin bands splitting exhibit alternating signs across the gamma point.\cite{Mazin2}

\begin{figure*}
\centering
\includegraphics[width=16. cm]{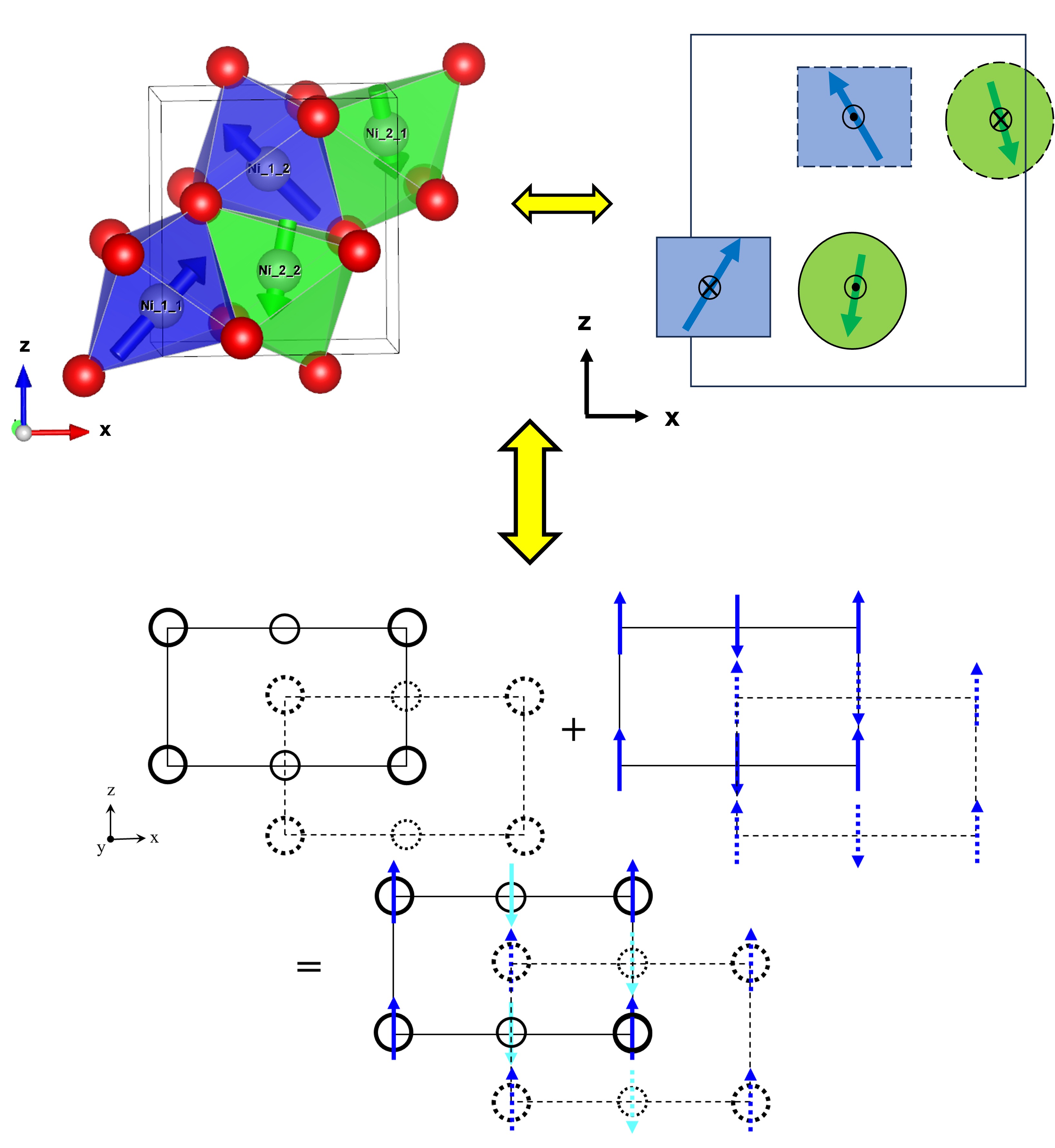} \vspace{-2mm}
\caption{\textbf{Top panel}- The two Ni atom spins belonging to the same type are related by mirror symmetry along ac plane and bc plane. The corresponding “octahedra” are equivalent via this mirror symmetry operation: (m$_{x}$, m$_{y}$, m$_{z}$)(-m$_{x}$, -m$_{y}$, m$_{z}$). Ni spins of different types (octahedra of different colors) are not symmetry equivalent. The entire unit cell is fully compensated along x, y directions, but can allow net magnetization along z direction due to the structural variation. \textbf{Bottom panel} – A simplified collinear-spin version of the magnetic state in NiSi. Two different silicon ion coordinations of Ni ions and Ni spins are shown with two different black circles and blue arrows, respectively. Crystallographic lattice of AB-stacked rectangular layers is centrosymmetric and the magnetic lattice with collinear spins is \textbf{PT}-symmetric, but the combination of the crystallographic lattice and the magnetic lattice has broken \textbf{PT} symmetry. Both the real magnetic state of NiSi and the cartoon in the bottom panel correspond to the magnetic point group of m'm'm with non-zero net magnetic moment along z of single layer.} \vspace{-4mm}
\end{figure*}

\begin{figure*}
\centering
\includegraphics[width=16. cm]{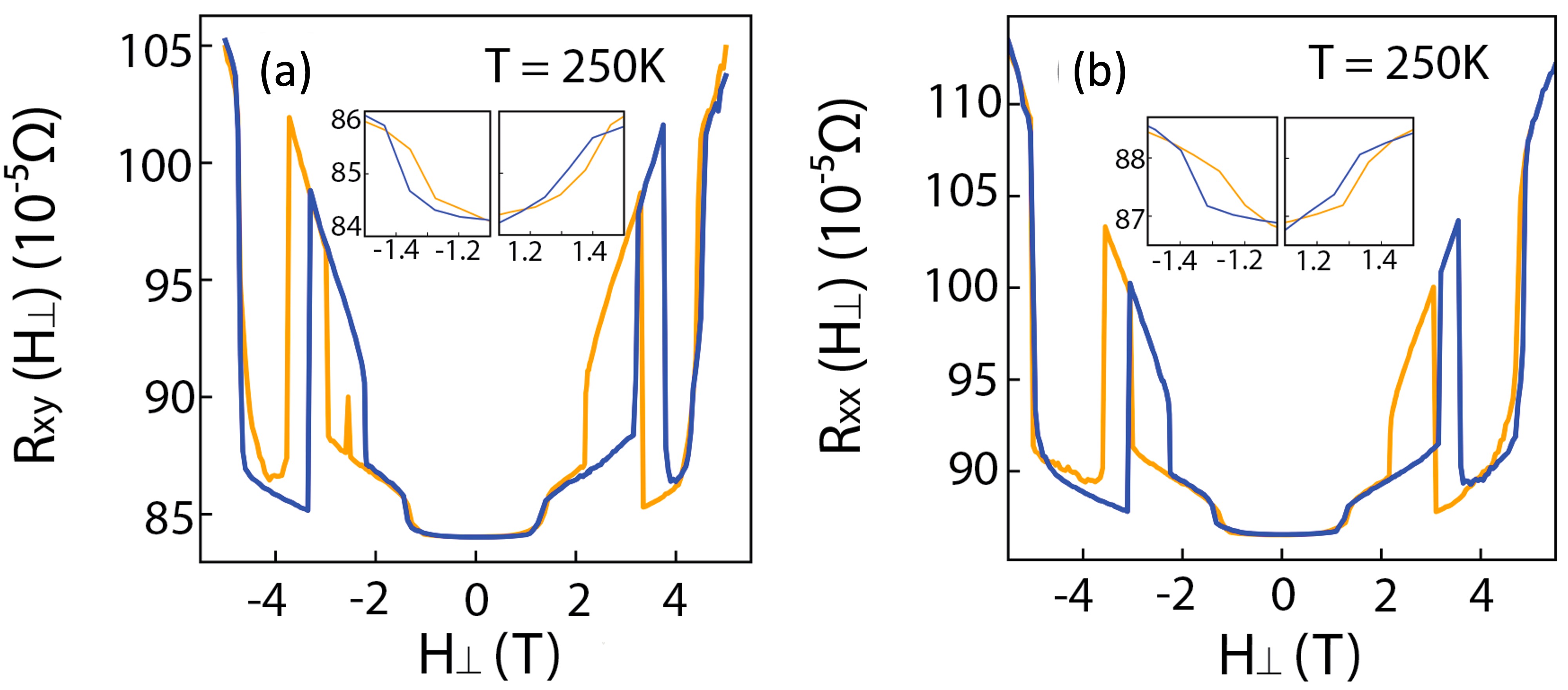} \vspace{-2mm}
\caption{\textbf{Plots of Hall resistance, R$_{xy}$, and linear resistance, R$_{xx}$, as a function of magnetic field applied along z-axis.} Strong asymmetries are observed in both R$_{xy}$ and R$_{xx}$ data. (Adapted from ref. [36])} \vspace{-4mm}
\end{figure*}

Another criterion in Cheong’s proposal links the altermagnetic property to the anomalous Hall effect (AHE) of odd order.\cite{Du,Smejkal3,Amar} The altermagnets must exhibit anomalous Hall resistance R$_{xy}$ either due to the existing ferromagnetic moments (in the so-called type-I), or magnetic moment induced by applied electric currents (in the so-called type-II). Since the anomalous Hall resistance is dependent on the band splitting of spin sub-lattices in AFM compounds, such experimental observations can account for the theoretically demonstrated asymmetric spin-split bands in altermagnetic systems.\cite{Yu} 

From experimental perspective, identifying altermagnetic material has been a challenge. But the experimentally observable criteria, envisioned in Cheong’s proposal e.g. crystalline asymmetry and AHE signatures, can act as guide in this pursuit. Among the two types of altermagnets, the type-I with net non-zero magnetic moment can be a more realistic proposition in AFM compounds with non-collinear spin configuration. They often break the \textbf{PT} symmetry and linked to the magnetic point group of the antiferromagnetic compound.\cite{Hayami,Sang,Turek} For example, any magnetic state with m'm'm can have a non-zero magnetic moment as well as a non-zero magnetic octupole. Uniaxial stress along xy or yx in 4'/mm'm group can induce a net magnetic moment along the z-axis. Similarly, uniaxial stress along x or y on m'm'm can induce an additional magnetic moment along the z direction. This induced magnetic moment can be additive or subtractive to the existing net magnetic moment, depending on the compressive or tensile stress. Any magnetic point group (MPG) belonging to the ferromagnetic point group can have non-zero net magnetization. Also, any MPG having magnetic octupole can have off-diagonal piezomagnetism,\cite{Aoyama,Bhowal,Hayami} thus causing non-linear anomalous Hall effect. Similar argument is theoretically proposed in the case of magnetic quadrupole moment.\cite{Hayami2} Nonetheless, it is established that broken \textbf{PT} symmetry can lead to the anomalous Hall effect.\cite{Smejkal3,Sang,Du} Thus, the two conditions outlined in Cheong's proposal to realize an altermagnet are intricately connected. 

In this perspective, we discuss several antiferromagnetic candidate materials with non-collinear spins that can host either type-I or type-II altermagnetic state. The discussion is divided into two parts: first part is focused on NiSi where we demonstrate the broken \textbf{PT} symmetry and non-linear AHE. In the second part, we describe the possibility of broken \textbf{PT} symmetry and non-linear AHE in antiferromagnetic candidate compounds, based on the magnetic point group and available Hall data in the literatures. Accordingly, we try to classify them in the two altermagnetic categories. The presented discussion is by no means a complete description of altermagnetism in AFM compound with non-collinear spins. The perspective is meant to generate a debate on this topic with the goal of finding many more such compounds in future. 

\begin{figure*}
\centering
\includegraphics[width=16. cm]{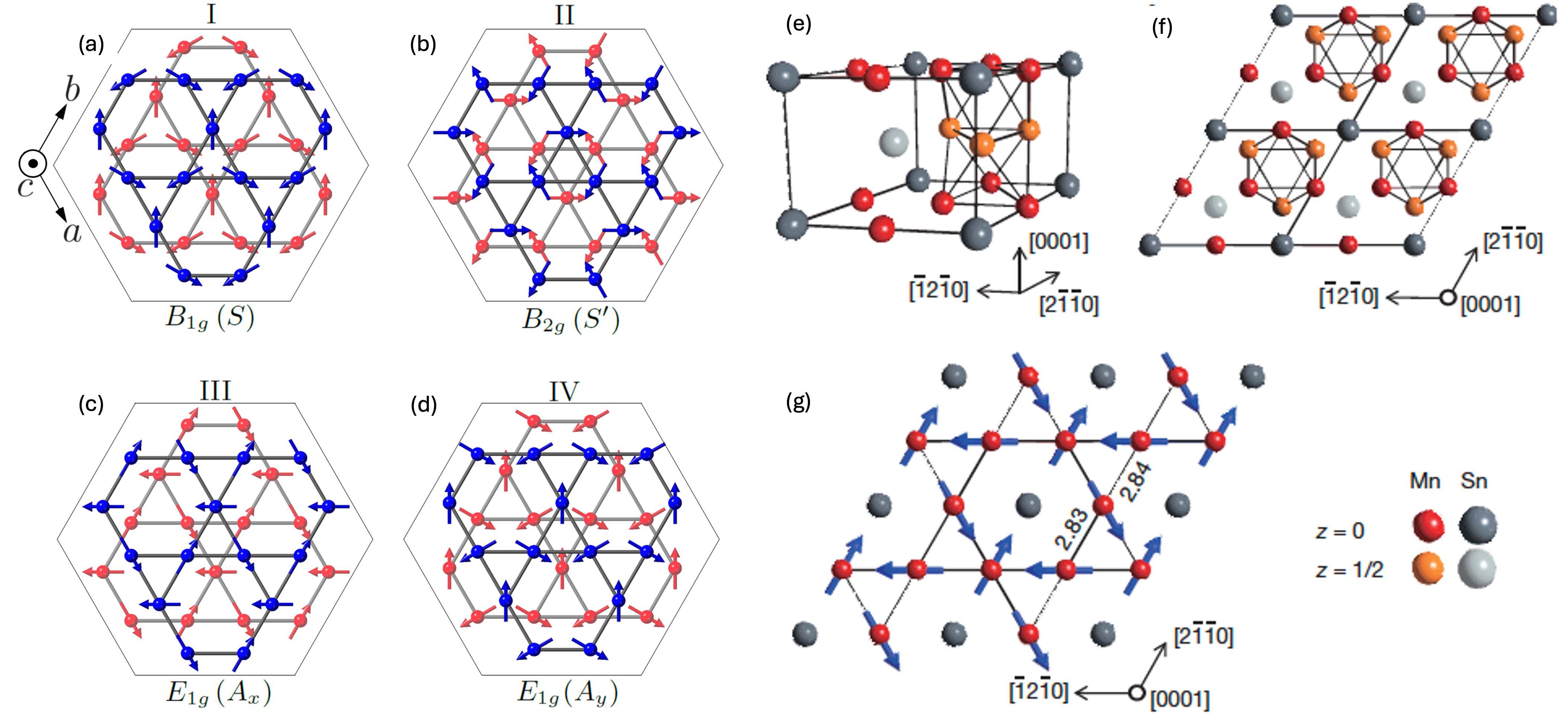} \vspace{-2mm}
\caption{\textbf{Magnetic structure of Mn$_{3}$Ge and Mn$_{3}$Sn.} (a-d) Irreducible representational analysis of neutron scattering data on Mn$_{3}$Ge by Soh et al. suggests four possible magnetic structures. They suggest that Mn ions occupy the Wyckoff position of 6$h$ with $z$ = 1/4 and 3/4. The structures shown transform according to the irreducible representations of the D$_{6h}$ point group. The symmetry label of the irreducible representation is given, together with the labels for the order parameters (in parentheses). The best fit to experimental data is obtained for the spin configuration shown in fig. (d). (Adapted from reference [49]). (e-g) Magnetic structure of Mn$_{3}$Sn. Unlike Mn$_{3}$Ge, magnetic structure in Mn$_{3}$Sn is best described by the inverse triangular spin structure shown in fig. (g). In Mn$_{3}$Sn, Mn ions are in z = 0 and z = ½ planes. (Adapted from reference [38]).} \vspace{-4mm}
\end{figure*}

\textbf{Altermagnetic state in NiSi}

Nickel monosilicide (NiSi) can be the first realistic altermagnet with non-collinear spin configuration. Nickel silicides are important electronic materials that have been used as contacts for field effect transistors, as interconnects, and in nanoelectronic devices.\cite{Lavoi} Although Ni is an elemental ferromagnet, no magnetic order was ever found in the Ni-Si compounds. Recently, neutron scattering measurements on single crystal NiSi revealed the antiferromagnetic ground state with very high Neel temperature, $T_N$ $\geq$ 700 K.\cite{Ghosh}But the magnetic unit cell is not fully compensated—there is a small ferromagnetic moment, accompanying the AFM order parameter. Also, it exhibits strong signatures of higher order AHE in the AFM plane. Thus, it can be a type-I altermagnet, subjecting to the fulfillment of \textbf{PT} violation criteria. 

NiSi crystallizes in the zinc blend-type orthorhombic structure, space group Pnma, with lattice parameters of a = 5.178 $\AA$, b = 3.331 $\AA$ and c = 5.162 $\AA$. In the unit cell, four Ni ions occupy the wyckoff positions. Each Ni ion has two nearest Si ions. However, as we can see in Fig. 1, the Si ions surrounding the Ni ions form two different types of octahedra, thus causing different local chemical environment to the latter. Spin structure of correlated Ni ions is discussed in detail in ref. [36]. The magnetic space group belongs to P2$_1$ (No. 4.7) group with a transformed basis set of (x, z, -y: ¼, 0, 0), with respect to (x, y, z; 0, 0, 0) of the parent chemical structure (Pnma). Most importantly, the spin directions and the associated magnetic moments on two Ni sites are different: while the Ni ion surrounded by the blue octahedron has spin components given by (m$_{x}$, m$_{y}$, m$_{z}$), the spin direction at other Ni ion (surrounded by green octahedra) is described by (-m$_{x}$, -m$_{y}$, m$_{z}$), with compensated $x$ and $y$ components and non-compensated $z$ component. The ordered moments at the two different Ni sites are 1.58(06) $\mu_B$ (blue octahedron) and 1.19(04) $\mu_B$ (green octahedron), respectively. Therefore, the magnetic unit cell is non-centrosymmetric and allows for an overall net magnetic moment along the z-axis. Now, if we look at the ligand coordination around Ni ions in Fig. 1, then it becomes apparent that it violates the inversion symmetry. While the same color octahedrons are related by the mirror symmetry, the octahedrons of different color are not related by the inversion symmetry. A set of rotation operations would be needed to go from one Ni site to another. These two properties, the violation of parity or inversion symmetry in conjunction with the requirement of rotational operation to map one magnetic sub-lattice to another, fulfill the key criterion for NiSi to be classified as type-I altermagnet.

\begin{figure*}
\centering
\includegraphics[width=16. cm]{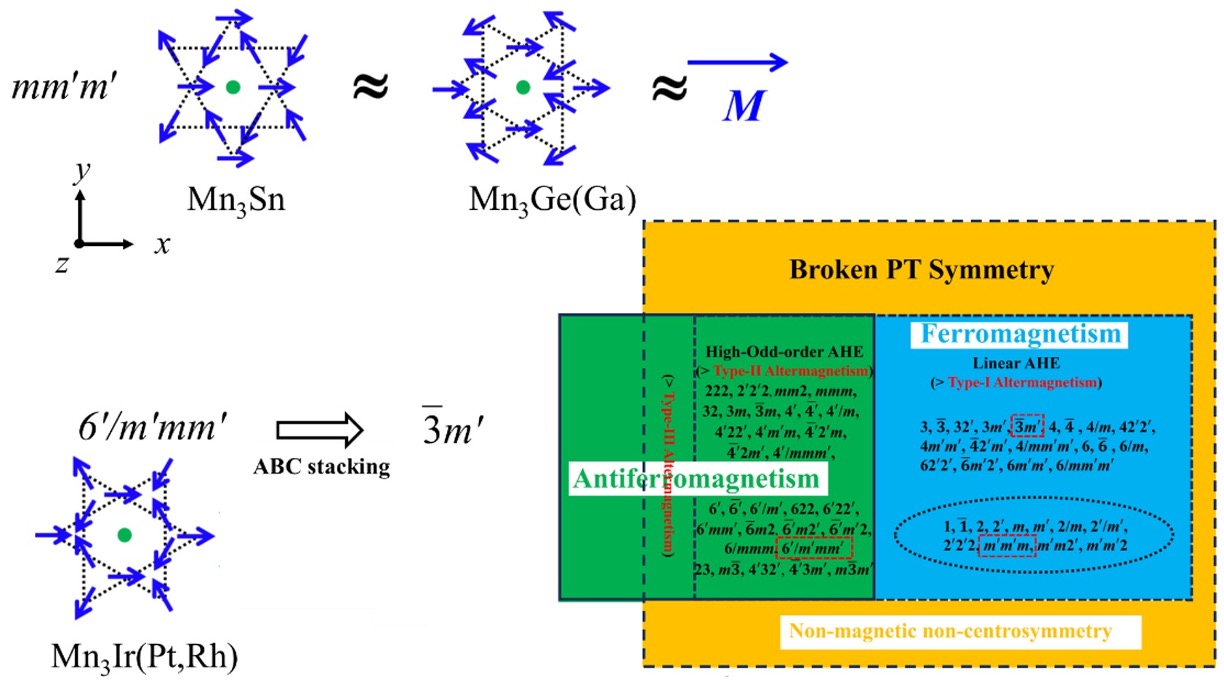} \vspace{-2mm}
\caption{\textbf{Magnetic space group and point group symmetry in Mn$_{3}$$X$, $X$ = Sn, Ge, Ir, Pt.} As shown in the figures, each of these compounds violate \textbf{PT} symmetry. (Partially adapted from reference [29]).} \vspace{-2mm}
\end{figure*}

Besides the asymmetric crystalline arrangement around magnetic ion, the report by Cheong et al. suggests that an altermagnet will also exhibit non-linear AHE for field application along the magnetic order parameter. Hall measurements on NiSi single crystal indeed reveal highly unusual magneto-electronic hysteresis for magnetic field application along the z-axis, perpendicular to the x-y plane, see Fig. 2. Both the Hall resistance R$_{xy}$ and the linear resistance R$_{xx}$ exhibit pronounced asymmetry in field sweep directions, resulting in hysteresis type patterns across $H$ = 0 T.\cite{Ghosh} The asymmetricity in Hall resistances become more pronounced at higher temperature (near and above room temperature), thus making it useful for practical applications in spintronics. However, Hall measurements at higher temperatures (at or above room temperature) are needed to verify moderate or low switching fields that can be practically applicable for spintronics device designs. As far as the underlying mechanism behind the magneto-electronic hysteresis is concerned, previously we had showed that the asymmetricity in R$_{xy}$ and R$_{xx}$ could be arising due to the rearrangement of magnetic spins along the crystallographic directions.\cite{Ghosh} However, NiSi violates \textbf{PT}-symmetry. Therefore, the non-linear AHE in NiSi may have contributions from the band splitting, as argued by theoretical researchers to realize an altermagentic state.\cite{Mazin,Smejkal,Smejkal2,Mazin2,Mccarthy} So, detailed theoretical research works are needed to understand the role of band splitting to the asymmetric Hall resistance in NiSi. 

\textbf{Other potential antiferromagnetic compounds with non-collinear spins for altermagnetism}

Besides NiSi, there are several several antiferromagnetic compounds with non-collinear spin configurations that can potentially host altermagnetic state. We focus on two sets of criteria for type I and type II classifications: a non-compensated magnetic unit cell plus anomalous Hall effect or a fully compensated antiferromagnetic state with anomalous Hall effect (AHE). 

\textbf{1. Mn$_{3}$$X$, $X$ = Sn, Ge, Ga}:

Two compounds of immediate interest are Mn$_{3}$Sn and Mn$_{3}$Ge.\cite{Naoki,Nakatsuji,XLi,Parkins,Balents,Yang,Higo,Tomiyoshi,Sung,Kiyohara,Nayak,Chen2,Soh} Both compounds crystallize in hexagonal Ni$_{3}$Sn-type structure with space group P6$_{3}$/mmc. The face-sharing octahedra of Mn atoms form a twisted triangular lattice tube along the c-axis. There are two Mn sites in z = 0 and z = 1/2 planes (in the case of Mn$_{3}$Ge, some reports suggest that Mn occupy the Wyckoff positions of z = 1/4 and 3/4), see Fig. 3. The compounds exhibits high temperature antiferromagnetism with $T_N$ = 420 K (Mn$_{3}$Sn) and $T_N$ = 380 K (Mn$_{3}$Ge). The Mn sublattices form slightly distorted kagome structure in the x-y plane. The spin configuration of Mn ions in Mn$_{3}$Sn is described by the inverse triangular structure, which results in small uncompensated ferromagnetic moment per atom in both compounds, see Fig. 3. Mn$_{3}$Ge manifests different spin correlation between Mn ions. Detailed analysis of symmetry allowed magnetic structure in Mn$_{3}$Ge is provided by Soh et al.\cite{Soh} The irreducible representations result in four magnetic structures of the D$_{6h}$ point group. The sixth-order hexagonal anisotropy splits the ground state degeneracy, resulting into an unusual magnetic ground state. Similar argument is applicable to the magnetic structure analysis of Mn$_{3}$Sn. Small uncompensated ferromagnetic moment due to the canting of moments is also reported in Mn$_{3}$Ge.

The chemical structure of Mn$_{3}$$X$ ($X$ = Sn, Ge) is such that it violates inversion symmetry, which is one of the requirements for altermagnetism. They belong to the magnetic point group of mm'm'. It is schematically depicted in Fig. 4. The small ferromagnetic component causes anomalous Hall effect in these materials, see Fig. 5. There are evidence of higher order AHE in these materials. There are some characteristic similarities in the Hall conductivity data in both Sn and Ge analogues. They both exhibit very large Hall conductivity, $\sigma$$_{xz}$ and $\sigma$$_{yz}$, for field application along c-axis. In fact, the Hall resistivity changes sign with the reversal of magnetic field, which was attributed to the rotation of staggered moments of the non-collinear spin configuration.\cite{Nakatsuji} It suggests the direct role of AFM order in the AHE. Both properties, the breaking of inversion symmetry and the unconventional AHE, are in accordance with the Cheong's proposal of manifesting altermagnetism. So, these compounds are potential candidates for type-I altermagnetism. 

\begin{figure*}
\centering
\includegraphics[width=16. cm]{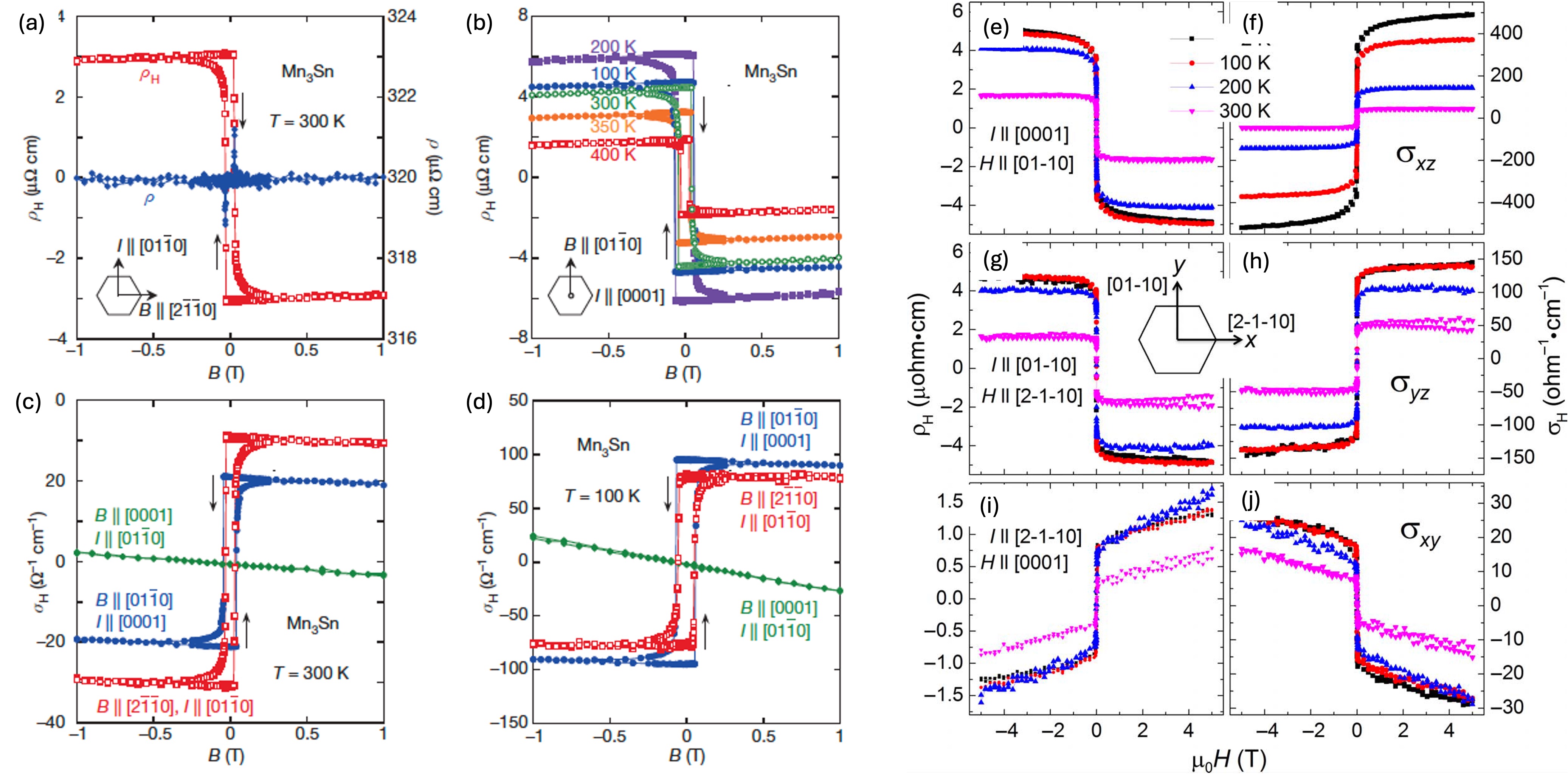} \vspace{-2mm}
\caption{\textbf{Magnetic field dependence of anomalous Hall effect in Mn$_{3}$Sn and Mn$_{3}$Ge}. (a-b) Hall (left axis) and longitudinal (right axis) resistivity as a function of magnetic field at different temperatures. Field application direction is shown in the inset. (c-d) Hall conductivity $\sigma_H$ as a function of field at $T$ = 300 K (fig. c) and 100 K (fig. d) for different combinations of current and field application directions. (e-j) Nonlinear Hall resistivity and AHE in Mn$_{3}$Ge. (Adapted from ref. [38] and [47], respectively).} \vspace{-2mm}
\end{figure*}

Another compound in this series is Mn$_{3}$Ga. Depending on the heat treatment during the synthesis process, Mn$_{3}$Ga can crystallizes in tetragonal, hexagonal and face-centered cubic structures.\cite{Liu,Zhang} Among them, the hexagonal structure is shown to exhibit similar inverse triangular spin structure at high temperature, see Fig. 6. At $T \sim$ 170 K, the compound undergoes subtle structural transition to the orthorhombic phase, causing large deviation in spin structure. It is also known to exhibit large AHE, owed to the non-collinear spin arrangement of the antiferromagnetic origin. Arguably, the Hall resistivity at low temperature is of topological origin.\cite{Liu}

\textbf{2. Mn$_{3}$X, $X$ = Rh, Ir, Pt}: 

\begin{figure*}
\centering
\includegraphics[width=16. cm]{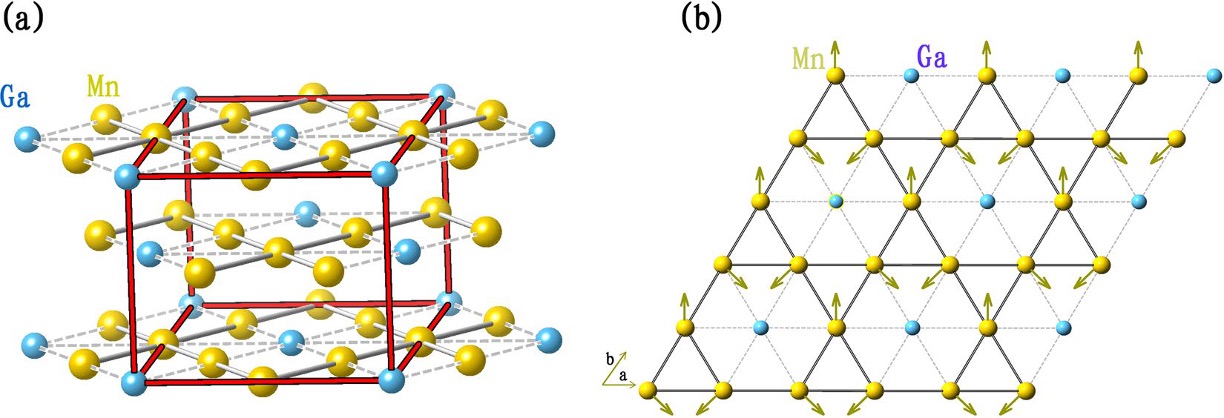} \vspace{-2mm}
\caption{\textbf{Chemical and magnetic structure in Mn$_{3}$Ga.} The spin configuration of Mn ions is similar to that of Mn$_{3}$Ge. (Adapted from reference [50]).} \vspace{-4mm}
\end{figure*}

\begin{figure}
\centering
\includegraphics[width=8.7 cm]{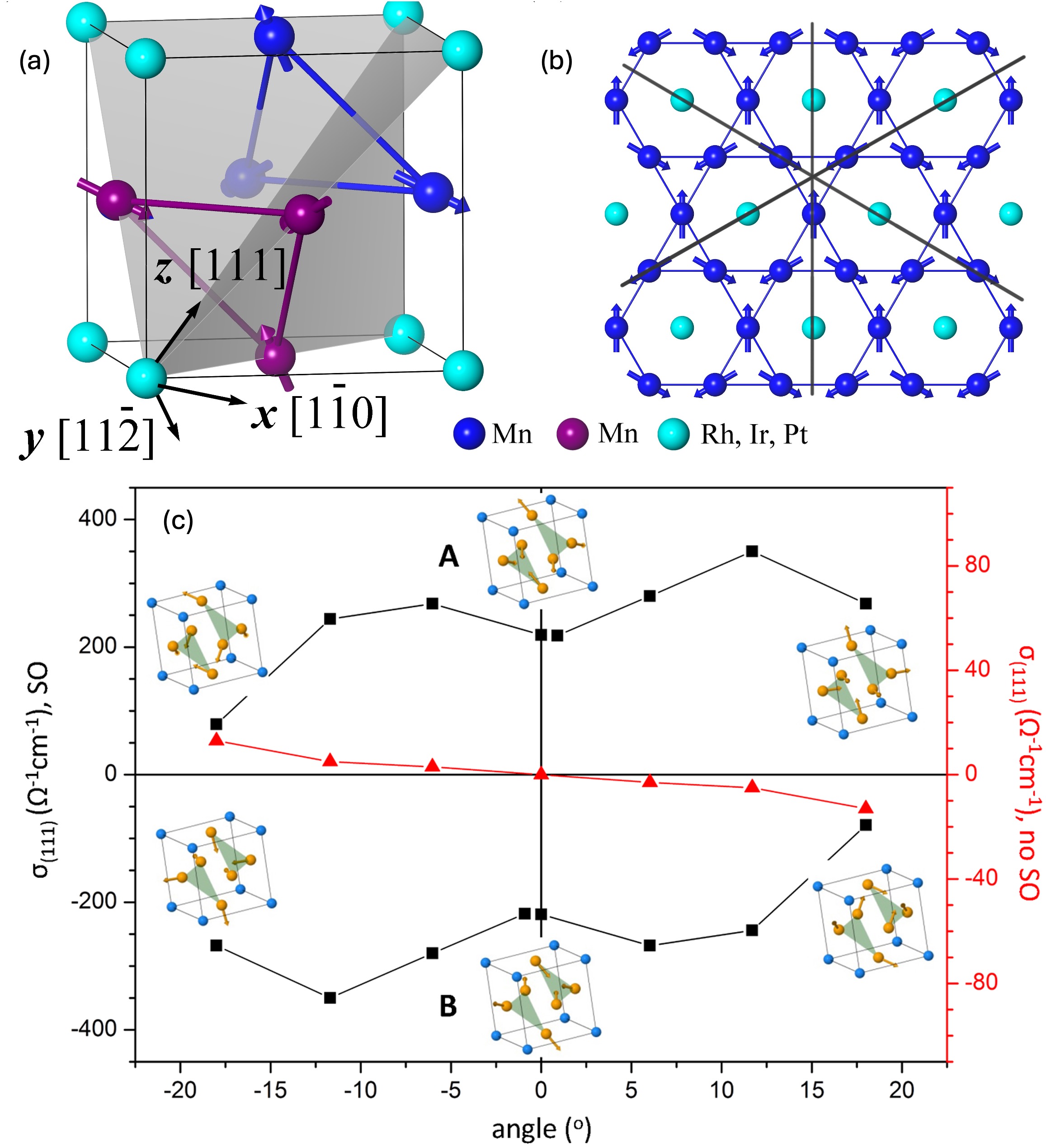} \vspace{-4mm}
\caption{\textbf{Magnetic structure and anomalous Hall effect in Mn$_{3}$X (X = Rh, Ir, and Pt) compound.} (a-b) Mn$_{3}$X (X = Rh, Ir and Pt) has an FCC lattice. Three mirror planes are shown in gray. The Mn sublattice forms a Kagome-type configuration. The projections of three mirror planes are indicated by black lines. Here, the Kagome plane is set as the xy plane and the plane normal as the z axis. A mirror refection combined with a time-reversal symmetry operation preserves the magnetic lattice. However, the alternation of two Mn kagome planes with opposite magnetic structures would require rotation operation of ligands coordination to map the magnetic lattice. (c) Theoretically predicted anomalous Hall conductivity (AHC) in different phases of Mn$_{3}$Ir. The plot shows AHC as a function of Mn tile angle with (black squares) and without (red triangles) spin-orbit coupling. (Adapted from references [51] and [53]).} \vspace{-8mm}
\end{figure}

\begin{figure*}
\centering
\includegraphics[width=16. cm]{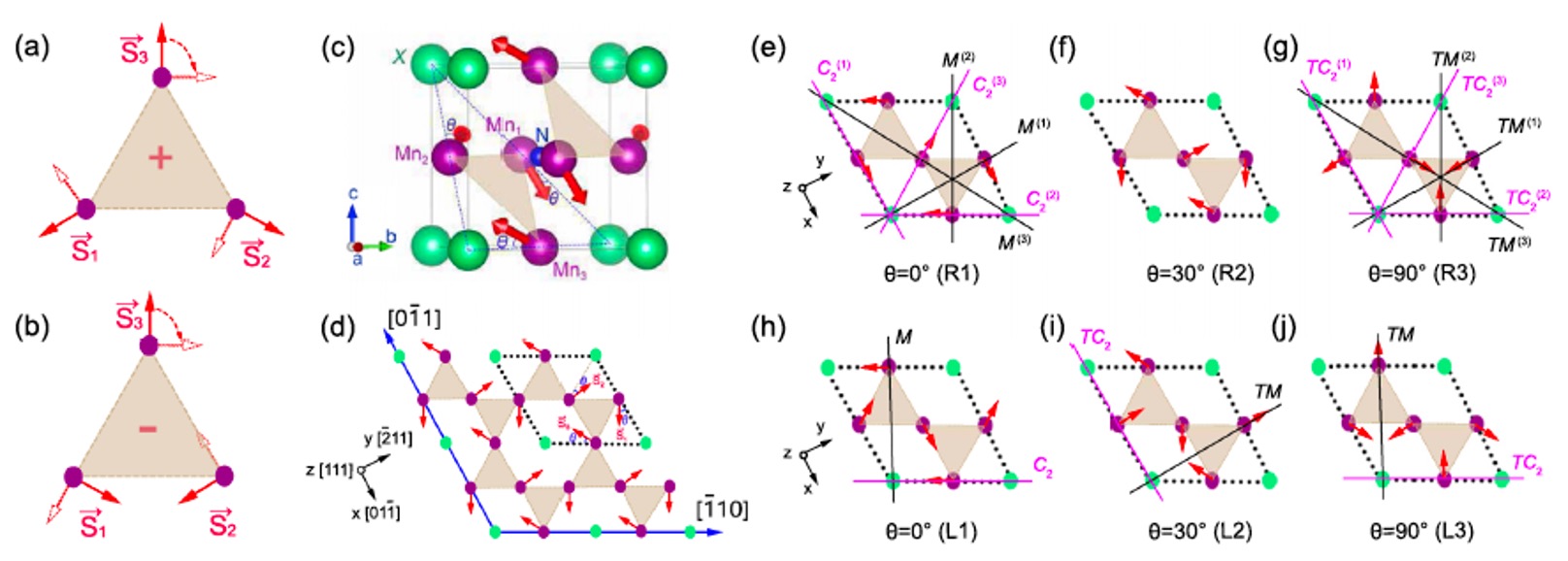} \vspace{-2mm}
\caption{\textbf{Spin chirality and spin structure in Mn$_{3}$$X$N.} (a-b) Right and left-handed spin chirality in Mn$_{3}$$X$N. The open arrows indicate the clockwise rotation of spin with a uniform angle, which results in a different spin configuration with the same spin chirality. (c-d) The crystal and magnetic structures of Mn$_{3}$$X$N. Mn moments are aligned within (111) plane. The dotted lines mark the two-dimensional unit cell. (e-g) The R1, R2, and R3 phases with the right-handed spin chirality. There are one three-fold rotation axis and three mirror planes in the R1 phase. (h-j) The L1, L2, and L3 phases with the left-handed spin chirality. There are one two-fold rotation axis (C2) and one mirror plane (M) in the L1 phase; the time-reversal symmetry T is combined with two-fold rotation and mirror symmetries in both the L2 and the L3 phases. (Adapted from ref. [60]).} \vspace{-4mm}
\end{figure*}

\begin{figure*}
\centering
\includegraphics[width=16. cm]{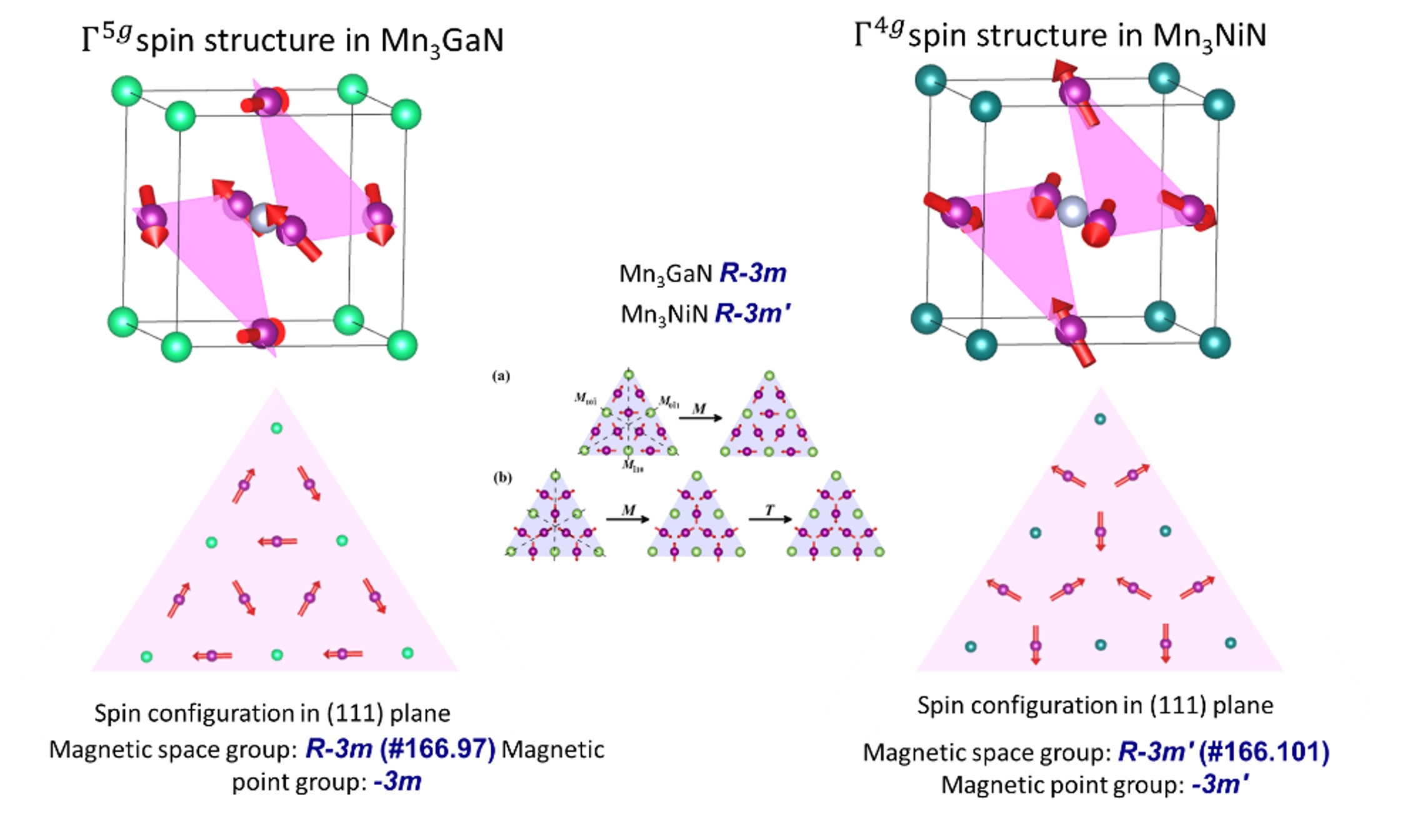} \vspace{-2mm}
\caption{\textbf{Magnetic point group in Mn$_{3}$$X$N.} Examples of magnetic point group in Mn$_{3}$$X$N compounds. Most of them exhibit -3$m$ or -3$m$' that violate \textbf{PT} symmetry. (Partially adapted from ref. [65]).} \vspace{-4mm}
\end{figure*}

\begin{figure*}
\centering
\includegraphics[width=16. cm]{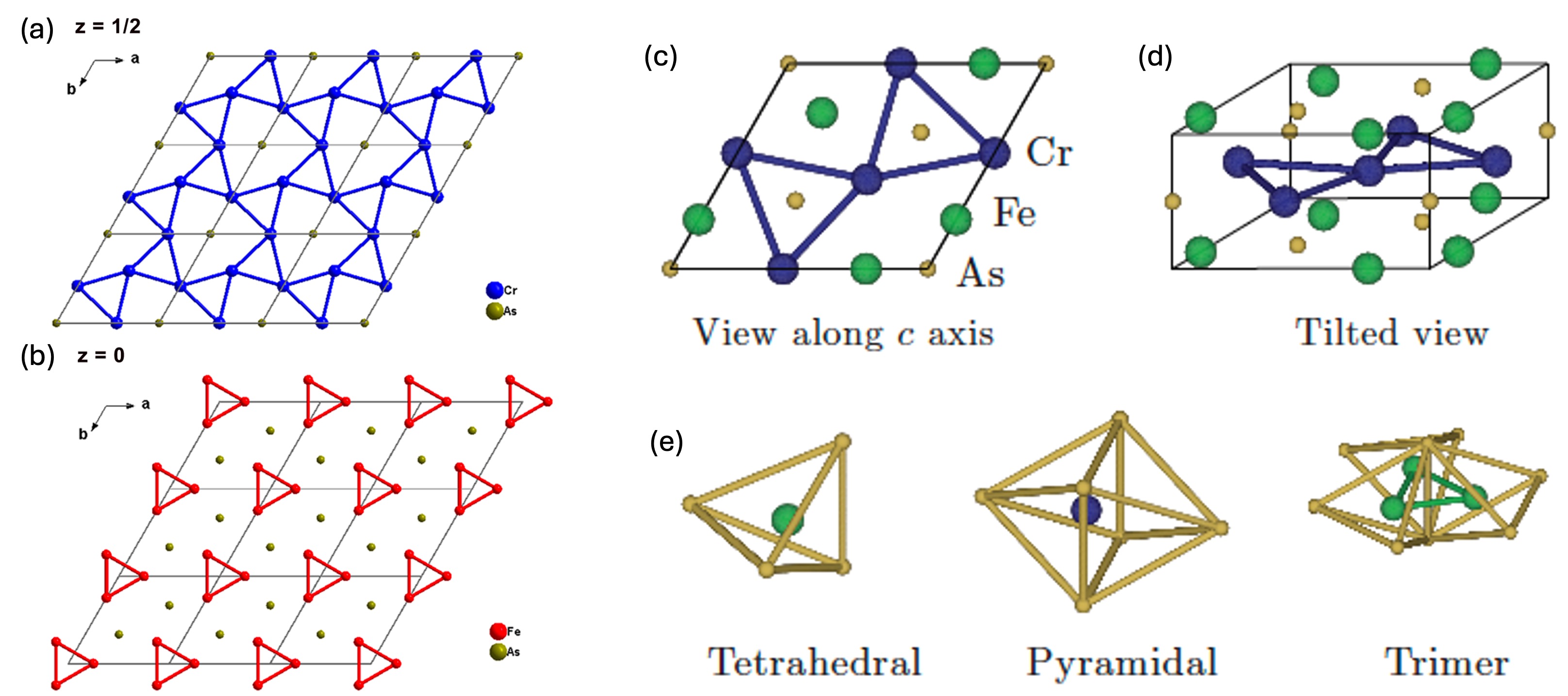} \vspace{-2mm}
\caption{\textbf{PT symmetry violation in non-collinear AFM compound FeCrAs.} (a-b) Chemical structure of FeCrAs. While the Fe ions form a triangular lattice of trimers, the Cr ions form a distorted kagome framework within the basal plane. These planes of the (a) Cr kagome framework and (b) Fe trimers stack alternately along the c axis with the As ions interspersed throughout both layers. (c-d) Unit cell viewed along the c-axis and tilted away by 70$^{o}$. (e) Ligand’s coordination of Fe, Cr and trimer. The Fe atom is tetrahedrally coordinated and the Cr atom is approximately octahedrally coordinated. The trimer is surrounded by three tetrahedra that share a common axis. (Adapted from references [68] and [70]).} \vspace{-4mm}
\end{figure*}

\begin{figure}
\centering
\includegraphics[width=9. cm]{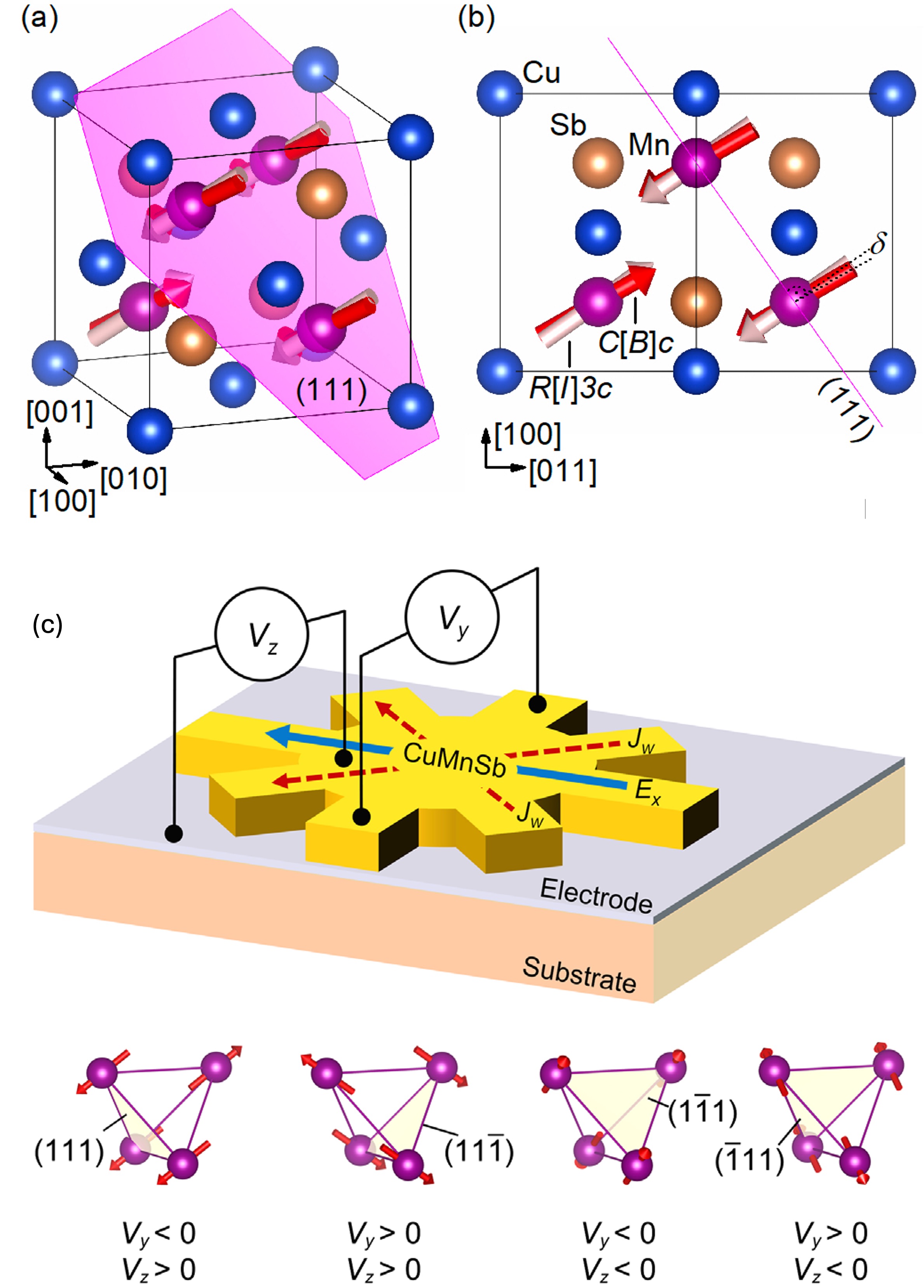} \vspace{-2mm}
\caption{\textbf{Magnetic structure and Hall resistivity in CuMnSb.} (a-b) Magnetic structure of CuMnSb is described by the magnetic space groups R[I]3c and C[B]c. Moments are perpendicular to the nuclear [111] plane (denoted in pink color) but canted away by small angle. (Adapted from ref. [40]) (c) Theoretically predicted anomalous Hall effect in CuMnSb. The signs of the Hall voltages $V_y$ and $V_z$ for different non-collinear AFM orders, resulting from the canting of moments in CuMnSb. (b) A spin-orbit torque device where the AFM orders of CuMnSb are switched using the antidampinglike SOT generated by the writing current $J_w$. (Adapted from references [74] and [76]).} \vspace{-4mm}
\end{figure}

Unlike the hexagonal structure manifested by Mn$_{3}$Sn or Mn$_{3}$Ge, Mn$_{3}$X, $X$ = Rh, Ir, Pt crystallize in the face centered cubic structure in the space group $Pm-3m$.\cite{Feng2,Chen,Tomeno,Iwaki,Cespedes,Felser,Huang,Arpaci,Zhang} Mn ions are located at the face-centered positions, forming triangular configuration in the [111] plane. All three compounds manifest very high Neel temperature: 960 K (Mn$_{3}$Ir), 850 K (Mn$_{3}$Rh), 470 K (Mn$_{3}$Pt). Mn moments, arranged in kagome lattice, are either pointing towards or away from the center of the triangle, see Fig. 7. Arguably, the cubic anisotropy is expected to cause out of plane distortion in Mn spin structure, which can result in small uncompensated ferromagnetic component. So, even though the chemical structure is significantly different from that of Mn$_{3}$Sn or Mn$_{3}$Ge, the overall magnetic characteristic of possessing small ferromagnetic component due to crystalline anisotropy induced out of plane canting of moment is somewhat similar. However, unlike in Mn$_{3}$Ge (or Sn analogue) where the Mn sublattices are stacked in -AB-AB- sequence along c-axis, the stacking is more complex in Mn$_{3}$Ir (or in Rh, Pt analogues). In the latter case, Mn sublattices are stacked in -ABC-ABC- sequence in [111] plane. Consequently, the ligands coordination varies from magnetic site-to-site. The magnetic point group of this series of AFM compounds belong to -3m' (see Fig. 4), which violates \textbf{PT} symmetry. Furthermore, the structural distortion in kagome lattice and the complex stacking profile makes the use of rotation operation necessary for lattice mapping.

\begin{figure*}
\centering
\includegraphics[width=16. cm]{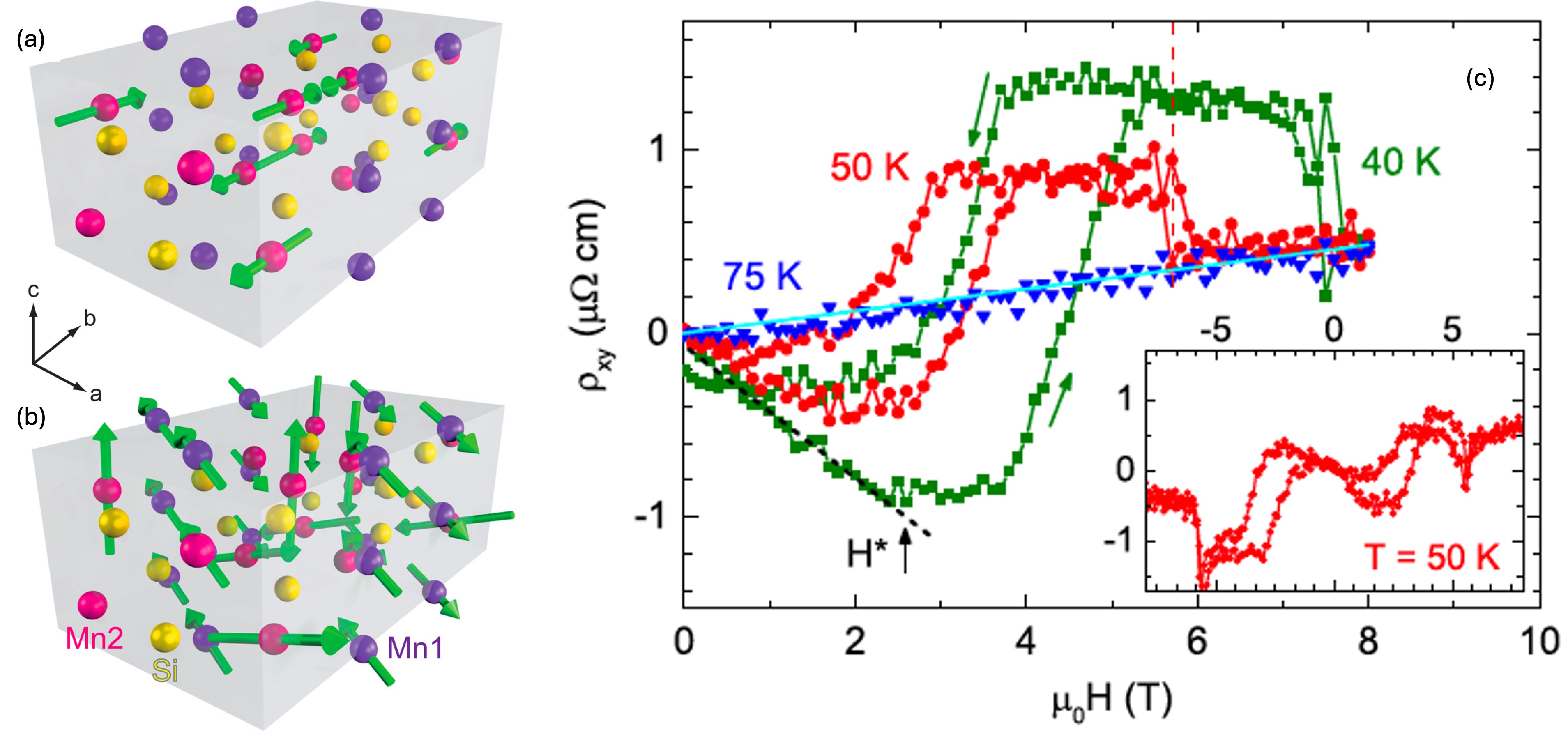} \vspace{-2mm}
\caption{\textbf{Non-collinear magnetic structure and Hall resistivity of higher order in Mn$_{5}$ Si$_{3}$.} (a-b) It crystallizes in the hexagonal space group $P6_3/mcm$ with two distinct crystallographic positions for the Mn atoms (sites Mn1 and Mn2). Magnetic structure of Mn$_{5}$ Si$_{3}$ undergoes a transition around $T \sim$ 62 K. At high temperature, Mn ions develop collinear AFM structure. Below that temperature, the Mn ions on two different sites exhibit diverse non-collinear arrangements with lower symmetry, violating PT symmetry.  (c) The transition in magnetic structure is also distinctly reflected by the metamagnetic transition in the Hall resistivity. In the collinear phase, $\rho$$_{xy}$ exhibits ordinary Hall effect (blue curve). (Adapted from reference [78]).} \vspace{-4mm}
\end{figure*}

These materials are also known to manifest strong anomalous Hall effect. Chen et al. have provided a foundational theoretical account of the AHE phenomena in Mn$_{3}$Ir, which is expected to be applicable to other compounds as well.\cite{Chen} For the first time, authors explicitly showed that antiferromagnets can have non-zero anomalous Hall conductivity due to the breaking of time reversal symmetry. Interestingly, the anomalous Hall effect is only detected along the [111] direction in Mn$_{3}$X, $X$ = Rh, Ir, Pt. Zhang et al. has explained this unusual phenomena.\cite{Zhang} Since there are three mirror planes in the cubic lattice, related by a three-fold rotation, that intersect the [111] axis, the only nonzero anomalous Hall conductivity component arises along the co-axis of these three planes, i.e. the [111] axis. Nevertheless, the combination of uncompensated ferromagnetic moment, \textbf{PT} symmetry breaking and the strong anomalous Hall effect makes them ideal candidate for the exploration of type-I altermagnetism. 

\textbf{3. Mn$_{3}$$X$N, $X$ = Ga, Zn, Ag, Ni}:

The set of compounds Mn$_{3}$$X$N render another potential platform to explore altermagnetism. Mn$_{3}$$X$N have the anti-perovskite structure and belong to the space group $Pm-3m$.\cite{XZ,Na,Harish,Hayami2,Beckert,Zhou2,Gurung,Pradhan} They crystallize in the cubic structure with Mn and X atoms occupying the face-centered and corner positions, respectively. N atoms are located at the center of the cube, see Fig. 8. Like Mn$_{3}$X, Mn spins are also arranged on kagome lattices in the [111] plane in Mn$_{3}$$X$N. The magnetic structure in Mn$_{3}$$X$N compounds are, arguably, unusually complex. While all compounds in this group exhibit non-collinear but co-planar spin configuration, they undergo temperature-dependent transition. As discussed by Xiaodong et al.,\cite{XZ} the spin structure undergoes transition from $\Gamma$$^{5g}$ configuration, with spins pointing along the diagonal of the face at $T \leq$ 166 K, to $\Gamma$$^{4g}$ configuration at $T \geq$ 166 K with spins pointing to the center of the triangle formed by three nearest Mn atoms. It is described in Fig. 8a. As shown in Fig. 8, there are one three-fold rotation axis $C_3$, which is along the [111] direction, three two-fold rotation axes ($C_2$$^{1}$, $C_2$$^{2}$ and $C_2$$^{3}$) and three mirror planes ($M$$^{1}$, $M$$^{2}$, $M$$^{3}$) in the R1 phase. Of which, only $C_3$ axis is preserved in the R2 phase. The magnetic point group (MPG) analysis of these compounds suggest that they either belong to -3m or -3m' group. Both MPGs violate the \textbf{PT} symmetry, see Fig. 9. The time-reversal symmetry $T$ has to be combined with a two-fold rotation and mirror symmetries in the R3 phase to map the neighboring Mn sites with current ligand's arrangements.

Mn$_{3}$$X$N compounds also exhibit very interesting anomalous Hall effect with possible origin in topological Berry curvature, vector product spin chirality or spin canting due to structural symmetry lowering induced by the large displacements of the manganese atoms away from high-symmetry positions.\cite{ Na,Hayami2,Beckert,Zhou2,Gurung,Pradhan}This is a widely studied problem. Theoretically, it is shown that the right and left handed vector chirality can lead to non-zero intrinsic Hall conductivity in $\sigma$$_{xy}$, $\sigma$$_{yz}$ and $\sigma$$_{zx}$, respectively, owing to the lowering of the symmetry. So, it shows that despite the compensated nature of antiferromagnetic order in Mn$_{3}$$X$N, the anomalous Hall effect can be observed due to the chiral magnetic structure. We note that the spin structure is non-collinear but coplanar. Therefore, the chirality due to the scalar triple product is zero. There are also reports of the non-zero Berry curvature resulting from the spin-orbital interaction between filled and empty bands that are evenly spread across the Fermi surface, which causes the AHE effect. However, in a contrasting report by Rimmler et al., the spin canting of Mn moments due to the structural symmetry lowering, giving rise to small uncompensated moment, is arguably linked to the anomalous Hall effect.\cite{Rimmler} 

Although the underlying mechanism behind the anomalous Hall effect in Mn$_{3}$$X$N compounds is debatable, but there is a complete agreement on the experimental observation itself. Depending on the nature of antiferromagnetic order, compensated or uncompensated, these systems can be of strong importance to study altermagnetism. If the AFM state is fully compensated, then they can be candidates for type-II altermagnetism. In the case of uncompensated state, it can host a type-I altermagnetic state.

\textbf{4. FeCrAs}:

FeCrAs can be a potential candidate to realize altermagnetism. This is an interesting material. Unlike the Mn-based compounds with localized moments, discussed so far, FeCrAs is arguably an itinerant system.\cite{Rou,Swainson} The compound manifests a unique interplay between chemical structure and local spin arrangement across the unit cell. FeCrAs has a hexagonal Fe$_{2}$P-type structure with the space group $P-62m$.\cite{Jin,Swainson,Plumb,Huddart,Lau} Interestingly, Fe magnetic moment is quenched and only Cr atoms contribute to antiferromagnetism. From the chemical structure perspective, Fe atoms form the triangular lattices of trimers, whereas the Cr ions form a distorted kagome framework within the basal plane. The planes of Fe trimers and Cr kagome framework stack alternately along the c-axis. As atoms form tetrahedral and octahedral coordination with Fe and Cr atoms, respectively. As we can see in Figure 10, the ligands coordination of magnetic ions varies from site-to-site. Therefore, rotation operation is required to map the lattice. From magnetic point group perspective, it belongs to m'm'2 group that breaks the \textbf{PT} symmetry.

According to the theoretical report by Rau et al.,\cite{Rou} Fe sublattice exhibits hidden spin liquid state, despite the strong magnetic order manifested by Cr. It is argued that the Cr ions develop spin density wave (SDW)-type order below $T_N$ $\sim$ 125 K. A recent report by Jin et al. have suggested that the Cr moments cant out of the plane at higher temperature ($T \geq$ 95 K). They performed detailed neutron scattering measurements and carried irreducible-representational analysis. The group concluded that the best fit to neutron data is obtained for the irreducible representation of $\Gamma_3$ + $\Gamma_6$, indicating large out of plane component of Cr moment. The spin canting occurs gradually with increasing temperature. The magnetic unit cell is 3x3 times the chemical unit cell with a total of nine Cr sites. Each of them has different moment, forming a SDW-type configuration. However, at higher temperature, moments start canting out of the plane. According to Jin et al., the process completes as the measurement temperature approaches the $T_N$ value.\cite{Jin} 

It is not yet understood if the system develops a net uncompensated moment along the c-axis, due to the out-of-plane canting of Cr moments. Nevertheless, the peculiar combination of magnetic and chemical structures makes FeCrAs an attractive candidate to explore altermagnetism. So far, there is no conclusive report of anomalous Hall effect of higher order in this compound. In a recent report, Lau et al. have demonstrated anisotropic electronic scattering in the system using Hall effect analysis.\cite{Lau} However, there is no report of non-linear Hall resistivity. Therefore, further study is needed.

\textbf{5. CuMnSb}:

CuMnSb is a half-Heusler allow, which manifests antiferromagnetism below $T_N$ = 55 K.\cite{Regnat,Jeong} This is probably the only Heusler compound with antiferromagnetic order. It crystallizes in the non-centrosymmetric C1$_{b}$ cubic structure under the space group $F-43m$ with one of the fcc sub-lattice being vacant. Magnetic properties of CuMnSb is dominated by the spin configuration of Mn ions. A comprehensive experimental investigation of magnetic structure in CuMnSb by Regnat et al. have revealed that Mn moments are canted away from the local $<$111$>$ axes by 11$^{o}$, see Fig. 11.\cite{Regnat} So, it exhibits a non-collinear spin structure, unlike the isostructural CuMnAs with collinear spins. However, the antiferromagnetic unit cell is fully compensated. The combination of non-centrosymmetric structure and canted spin configuration violates \textbf{PT}-symmetry. CuMnSb is also argued to exhibit non-linear anomalous Hall effect. Using ab-initio calculations Shao et al. showed that the anti-damping spin-orbit torques can switch the Neel vector, which in essence can produce a sizable Berry curvature dipole and hence the nonlinear AHE.\cite{Shao} However, experimental confirmation to this effect is currently lacking. Perhaps, future research works can unveil that. The combination of \textbf{PT}-symmetry violation and the non-linear AHE in CuMnSb fits aptly to Cheong's proposal of altermagnetism.

\textbf{6. Mn$_{5}$Si$_{3}$}:

Mn$_{5}$Si$_{3}$ is another itinerant antiferromagnetic system with non-collinear spin configuration of Mn ions.\cite{Biniskos,Surgers,Lander,Miina,Das,Kounta} Actually, the compound undergoes two successive AFM transitions at $T \sim$ 100 K and 60 K, respectively. While the Mn ions are mainly collinear in the high temperature AFM phase, it exhibits a non-collinear structure below $T \sim$ 60 K, see Fig. 12. Mn$_{5}$Si$_{3}$ is known to depict a peculiar inverted hysteresis loop. Crystallographically, it crystallizes in the hexagonal structure in the space group of $P6$$_{3}$/$mcm$ with two distinct Mn atoms (Mn$_{1}$ and Mn$_{2}$). At reduced temperature, it undergoes a structural transition to the orthorhombic phase with reduced symmetry in the space group $Ccmm$. In the orthorhombic phase, the Mn$_{2}$ site gets further divided into two more distinct sites. So, there is a total of four different Mn sites at $T \leq$ 60 K. Biniskos et al. argue that the magnetic structure in Mn$_{5}$Si$_{3}$ has monoclinic or possibly lower symmetry, even if they have similar chemical environment, at low temperature.\cite{Biniskos} So, the lattice cannot be mapped by simple translation or translation plus mirror operation. There is an intrinsic violation of \textbf{PT} symmetry in Mn$_{5}$Si$_{3}$. It is also known to exhibit anomalous Hall effect, which becomes very pronounced at low temperature. The AHE signal is further demonstrated to manifest a metamagnetic transition above a critical field $H$$^{*}$ in the single crystal sample.\cite{Surgers} The metamagnetic transition is very similar to that observed in NiSi crystal. Clearly, Mn$_{5}$Si$_{3}$ fulfills both criteria of Cheong's proposal for type-II altermagnetism. Further study in this regard can elucidate the intrinsic altermagnetism in this compound.

\textbf{Summary}

We have provided a comprehensive experimental outlook to realize both type-I and type-II altermagnets in magnetic materials with non-collinear spins in this perspective. There has been a lot of emphasis on the study of collinear spins antiferromagnet in the quest to find altermagnetic state in recent times. Antiferromagnetic systems with non-collinear spin configurations are ignored, despite the fact that they provide a highly versatile platform to study the interplay between magnetic and crystal symmetries, which in some cases lead to the violation of \textbf{PT}-symmetry. Arguably, the \textbf{PT}-symmetry violation is linked to the development of quadrupolar or octupolar moments that induce non-linear and anomalous Hall effect. In other words, a natural consequence to this novel manifestation arise in the form of nonlinear Hall effect with contribution to anomalous signal. So, the Cheong's proposal of realizing altermagnetism focuses on these two interlinked conditions. At a more fundamental level, the PT-symmetry breaking is associated to certain magnetic point groups (see Ref. [29] for detail) that are also used to study altermagnetism in AFM compounds with collinear spins. Additionally, there is no concrete theoretical argument, which forbids the non-collinear AFM compound from hosting an altermagnetic state. It is also worth mentioning that the concept of altermagnetism is still in its infancy and a lot of theoretical and experimental research works are needed to mature the understanding. Therefore, it is highly desirable that we explore both collinear and non-collinear spins antiferromagnetic compounds.

\begin{figure}
\centering
\includegraphics[width=8.7 cm]{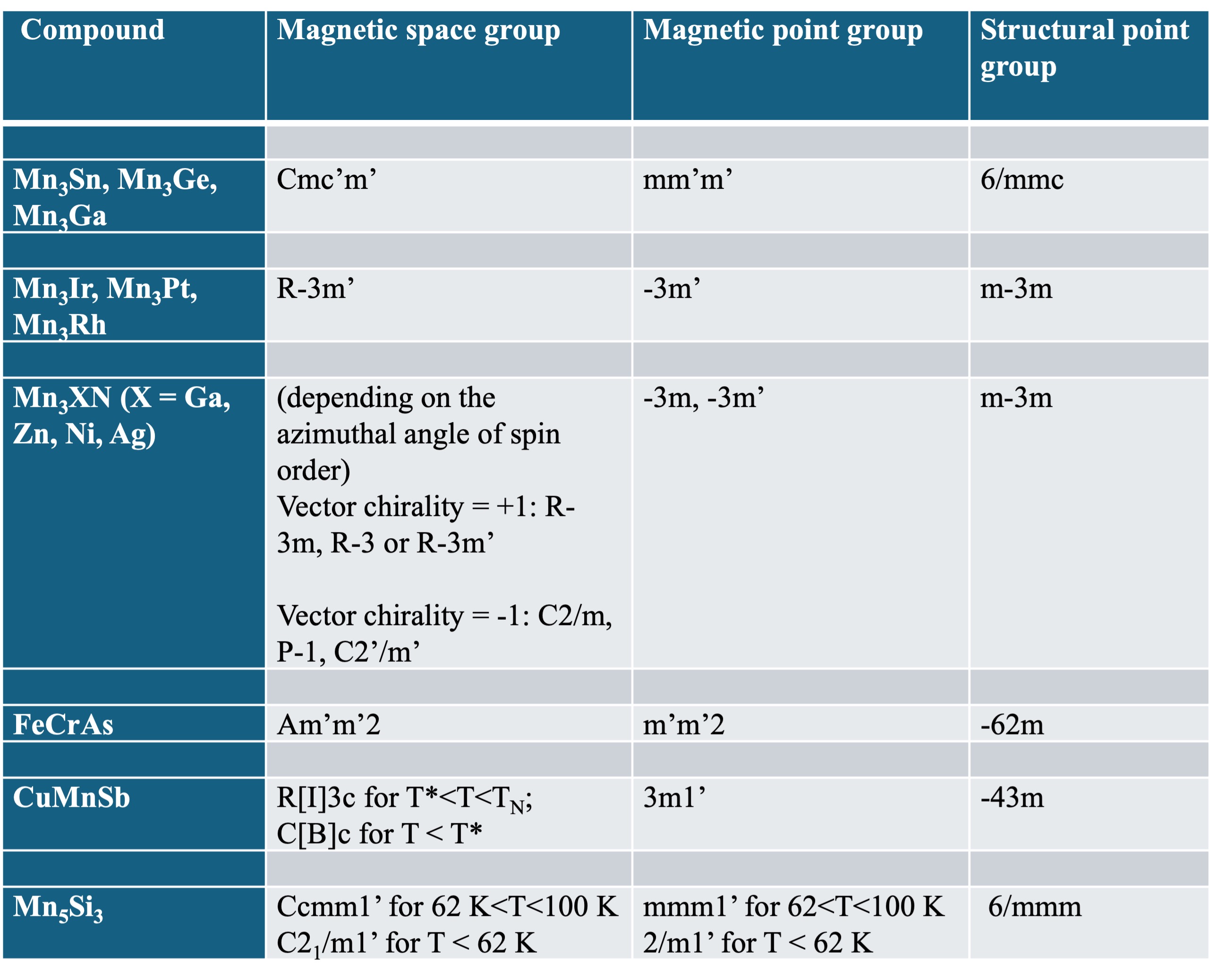} \vspace{-2mm}
\caption{\textbf{Figure table summarizing the magnetic space group and point group of the discussed compounds.} As discussed in the previous sections, the magnetic point group of each compound violate \textbf{PT}-symmetry. This is one of the key criteria for realizing altermagnetism in antiferromagnetic compounds with non-collinear spins.
.} \vspace{-4mm}
\end{figure}

This perspective is focused on non-collinear spins. In this effort, we have examined a series of potential candidate materials that may be suitable for further experimental and theoretical investigations of altermagnetism. We have used two key criteria in this regard: the \textbf{PT}-symmetry violation and the nonlinear Hall effect with finite AHE. As discussed in the previous paragraph, AHE may be a consequence of the quadrupolar or octupolar moments that also break the \textbf{PT}-symmetry. In Figure13, we have summarized the magnetic point groups of the compounds discussed in this perspective. All of them violate the \textbf{PT}-symmetry. In general, if the AFM compound has a net uncompensated ferromagnetic moment, then it is argued to exhibit anomalous Hall effect.\cite{Sang} Those compounds are classified under the type-I altermagnetims, subjected to the condition that they violate \textbf{PT}-symmetry. AFM materials with non-collinear spins and fully compensated magnetic unit cells that also violate \textbf{PT}-symmetry are proposed to be investigated for type-II altermagnetism, as outlined in Cheong's proposal. Although, we have discussed a list of materials in this regard, but there may be more AFM compounds with non-collinear spins that we were not able to explore. So, further study is needed. Finding direct connections between the theoretically predicted properties and the experimentally observable quantities will pave the way for new materials venues. Besides X-ray magnetic circular dichroism method, neutron scattering can be a powerful experimental technique to probe altermagnetic ground state in candidate materials.\cite{Maier}

Given the considerable interest in this emerging topic, further research works are highly desirable. Altermagnetic materials are also argued to be useful for spintronics applications.\cite{Sun} Therefore, this topic holds great interest for practical applications as well. However, in many cases, it derives from the unusual non-linear antiferromagnetic configuration of the host compound, which happens to violate the \textbf{PT}-symmetry.\cite{Yan,Jungwirth,Bai} NiSi or Mn$_{3}$Sn are apt examples to that. So, there are some intrinsic connections between the spintronic properties, non-collinear AFM with uncompensated moment and the PT-symmetry violation. But, it is not evident how an altermagnetic state can herald a new platform for the spintronics application at this time. Perhaps, future research will unveil that.

\textbf{Acknowledgements}

DKS acknowledges support from the U.S. Department of Energy, Office of Basic Energy Sciences under Grant No. DE-SC0014461. SWC was supported by the DOE under Grant No. DOE: DE-FG02-07ER46382.

\textbf{Conflict of Interest}

The authors declare no conflict of interest.

\textbf{Keywords}

Altermagnetism, Antiferromagnetic metal, Transition metal intermetallics, Spintronics.

\clearpage

\end{document}